\documentclass[aps,prb,reprint,amsmath,amssymb,groupedaddress]{revtex4-1}

\usepackage[colorlinks=true, allcolors=blue]{hyperref}
\usepackage{graphicx}
\usepackage{dcolumn}
\usepackage{bm}
\usepackage{color}
\usepackage{siunitx}
\usepackage{caption} 
\usepackage{subcaption}
\usepackage{todonotes}
\usepackage{threeparttable}
\usepackage{gensymb}
\usepackage{float}
\usepackage{soul}
\usepackage{xr}

\begin{document}

\title{First-principle study of multiple metastable charge ordering states in La$_{1/3}$Sr$_{2/3}$FeO$_3$}

\author{Nam Nguyen$^{1}$, Alex Taekyung Lee$^{2,3}$, Vijay Singh$^{4}$,  Anh T. Ngo$^{2,3}$, and Hyowon Park$^{1,2}$}

\affiliation{$^1$Department of Physics, University of Illinois at Chicago, Chicago, IL 60607, USA \\
$^2$Materials Science Division, Argonne National Laboratory, Argonne, IL, 60439, USA \\
$^3$Department of Chemical engineering, University of Illinois at Chicago, Chicago, IL 60608, USA \\
$^{4}$GITAM School of Science, Bangalore 561203, India 
}


\begin{abstract}
La doped SrFeO$_3$, La$_{1/3}$Sr$_{2/3}$FeO$_3$, exhibits a metal-to-insulator transition accompanied by both antiferromagnetic and charge ordering states along with the Fe-O bond disproportionation below a critical temperature near 200K. 
Unconventionally slow charge dynamics measured in this material near the critical temperature [Nature Communications, \textbf{9} 1799 (2018)] shows that its excited charge ordering states can exhibit novel electronic structures with nontrivial energy profiles.
Here, we reveal possible metastable states of charge ordering structures in La$_{1/3}$Sr$_{2/3}$FeO$_3$ using the first-principle and climbing image nudged elastic band methods.
In the strong correlation regime, La$_{1/3}$Sr$_{2/3}$FeO$_3$ is an antiferromagnetic insulator with a charge ordering state of the big-small-big pattern, consistent with the experimental measurement of this material at the low temperature.
As the correlation effect becomes weak, we find at least two possible metastable charge ordering states 
with the distinct Fe-O bond disproportionation. 
Remarkably, a ferroelectric metallic state emerges with the small energy barrier of $\sim$7meV, driven by a metastable CO state of the small-medium-big pattern.
The electronic structures of these metastable charge ordering states are noticeably different from those of the ground-state.
Our results can provide an insightful explanation to multiple metastable charge ordering states and the slow charge dynamics of this and related oxide materials.

\end{abstract}

\maketitle

\section{Introduction}

Charge ordering (CO) or charge density wave (CDW) is an intriguing material property driven by a spontaneous symmetry breaking of the periodicity in crystals. In strongly correlated materials, the charge degree of freedom is typically coupled to other degrees of freedom including spin, orbital, or lattice. While the origin of CDW can be purely electronic and the electronic correlation plays an important role,
it is often accompanied by structural distortions such as the bond-order or the Peierls transition, possibly leading to ferroelectricity.
Indeed, the combination of CDW, spin density wave (SDW), and the bond order has been proposed as the mechanism of ferroelectricity~\cite{Brink_2008,Ferroic_Orders1,LDA+DMFT_Ferroelectrics}.

La$_{1/3}$Sr$_{2/3}$FeO$_3$ (LSFO) is a transition metal oxide with a perovskite structure undergoing a weakly first-order transition at a temperature, $T$=200K, from a paramagnetic metallic state with the average valence state of Fe$^{3.67+}(d^{4.3})$ at a high temperature to an antiferromagnetic (AFM) insulating state with a CO sate of Fe$^{3+}(d^{5})$: Fe$^{5+}(d^{3})$=2:1 at a low temperature~\cite{EM1}.
Structural properties of LSFO with or without CO phases have been characterized experimentally using X-ray diffraction, neutron diffraction, and electron microscopy.
The studies of X-ray and neutron diffraction  \cite{Battle1, M&X&N1,M&X&N2,M&X&N3, M&X1,M&X1,XR_hole,MSandND} showed that bulk LSFO forms a rhombohedral structure in the space 
group of $R\bar{3}c$ 
(see Fig. \ref{1})
with the lattice constants $a=5.47$ \AA\: and $c=13.35$ \AA \: . 
A sign of the CDW spanning the periodicity of three Fe ions accompanied by SDW with a periodicity of six Fe ions was measured along the pseudocubic $[111]$ direction, but there was no clear evidence of structural distortions. Later, the electron microscopy study by Li \textit{et al.} \cite{EM2} revealed a structural distortions along the pseudocubic [111] direction in the real space upon the CDW transition. 
Finally, the neutron diffraction studies by Sabyasachi \textit{et al.} \cite{M&X&N1} and Yang \textit{et al.}\cite{M&X&N3} also showed a possibility of the meta-stable CO state due to multiple neutron peaks below the critical temperature.

Electronic properties of LSFO at the low-temperature CO phase haven been characterized by various experiments.
The study of optical spectroscopy by Ishikawa \textit{et al.} \cite{Optical1} showed the optical gap of LSFO was about 0.13 eV at low temperature. 
The studies of M\"{o}ssbauer spectrocopy \cite{M&X&N1,M&X&N2,M&X&N3,M&X1,MS,Battle1,MSandND,M1} captured two kinds of Fe ions with different hyperfine fields, confirming the charge disproportionation below the critical temperature. 
Recent ultrafast X-ray measurement in LSFO by Zhu \textit{et al} has shown that the noticeable slowdown occurs during the relaxation of CO near the critical temperature~\cite{Ultrafast}. They argued that the photoexcitation due to an ultrafast pump can drive  a ground state of La$_{1/3}$Sr$_{2/3}$FeO$_3$ into metastable states with different spin/charge orderings, which can be the origin of slowdown in the relaxation process. According to Yamamoto \textit{et al.}, \cite{Ultrafast_Yamamoto} these metastable or transient states are the CO in sequence of Fe$^{4+}$Fe$^{3+}$Fe$^{4+}$. However, the magnetic moments as well as the spin states, i.e. high spin (HS) or low spin (LS), of these Fe$^{4+}$ and Fe$^{3+}$ ions were unknown.  
In general, the slow dynamics of CO can be originated from the multiple meta-stable CO states accessible during the relaxation process.  

Unlike those various experimental characterizations,
theoretical studies of LSFO have been rather limited. The Hartree-Fock study by Matsuno \textit{et al}~\cite{hole} captured an energy gap of $0.14$ eV , which was in a good agreement with the experimental gap at low temperature. The first-principle study of density functional theory plus the Hubbard $U$ (DFT$+U$) by Zhu \textit{et al.}~\cite{Ultrafast} and Saha-Dasgupta \textit{et al.}~\cite{DFT+U} verified the presence of structural modulation or oxygen breathing distortions accompanied by CO of Fe ions 
in a sequence of 
Fe$^{3+}$Fe$^{5+}$Fe$^{3+}$.
They also found that another sequence of CO is possible, namely Fe$^{4+}$Fe$^{3+}$Fe$^{4+}$.
These CO states are strongly coupled to the spin states as the Fe ion with a larger charge state shows the high-spin state with the Fe-O bond elongation. Finally, the possibility of ferroelectricity in La$_{1/3}$Sr$_{2/3}$FeO$_{3}$ was pointed out by Park \textit{et al.} by rearranging the La/Sr layers~\cite{Ferro}.
Nevertheless, the effect of electronic correlations on the stability of CO states and the emergence of novel metastable states, such as ferroelectricity, which can be accessible from the photo-excitation experiment, have not been studied from first-principle. 

In this work, we study the effect of electron correlations on structural and electronic properties of LSFO having the strong charge-spin-lattice coupling by adopting the first-principle DFT+$U$ method. 
In particular, we explore possible meta-stable CO phases driven by a new pattern of structural distortions by adopting the climbing image nudged elastic band (CINEB) method along with DFT+$U$.
Remarkably, we find a new electronic phase in LSFO exhibiting the ferroelectricity driven by a small-medium-big CO pattern and a distinct Fe-O bond disproportionation with the small-medium-big magnetic moments.
This new meta-stable phase has almost the degenerate energy compared to previously known CO phases with a small energy barrier of $5\sim 7$ meV, implying the promising tunability of this material as a future electronic device.

\section{METHODS}

\subsection{First-principle calculation}

To perform the structural relaxation and the band structure calculations of LSFO, we adopt DFT+$U$~\cite{LDA+U} based on the projected-augmented wave (PAW) method~\cite{PAW} as implemented in the Vienna \textit{ab initio} simulation package (VASP)~\cite{vasp1,vasp2}. The exchange-correlation energy functional was treated using generalized gradient approximation (GGA) by adopting the Perdew-Burke-Ernzerhof (PBE) functional~\cite{PBE}. 
The cutoff energy for the plane-wave basis was used as 600 eV, and the Gamma-centered 8$\times$8$\times$2 $k-$point mesh~\cite{k-point} was used for all calculations. 
For structural relaxations, the Hellmann-Feynman force~\cite{Hellmann-Feynman} on each atom was set to be smaller than 0.01 eV/\AA. To treat the correlation effect of Fe $d$ orbitals, we impose the Hubbard $U$ and the Hund's coupling $J$ within DFT+U.

As noted from the previous study of Ref.~\onlinecite{Ultrafast}, two distinct CO structures 
(CO1 and CO3, see Fig. \ref{2})
can be obtained in LSFO by relaxing the crystal structure imposing different $U$ values 
on the Fe ions.
For the ground-state CO1 structure, we used $U$=5eV and $J$=1eV, 
while a distinct CO3 structure is obtained using $U$=3eV and $J$=0.6eV.
Both the crystal shape and ionic positions were relaxed during the structural relaxation, while the crystal volume of LSFO was fixed to 329.24 \AA$^{3}$.
To obtain a meta-stable CO phase (CO2, see Fig. \ref{2}), we adopt the CINEB method along with the DFT+$U$ using $U$=3.62eV (see Sec.\:\ref{sec:path}).
We also explore the effect of $U$ values ($J=0.2U$) on the stability of different CO phases (see Sec.\:\ref{sec:Udep}).

\subsection{Energy calculation along a structural path}

To obtain the minimum energy curve along a structural path and explore possible metastable structures,
we adopt the CINEB method along with DFT$+U$. The nudged elastic band (NEB) method is an efficient tool for finding the minimum energy path between two stable structures, i.e. a given initial (reactant) and final (product) state~\cite{MEP1,MEP2}. The CINEB method is a small modification of the NEB method without adding any significant computational method~\cite{CINEB}. 
The CINEB method yields a rigorous convergence of a saddle point, which has a maximum energy along the band but minimum in the other directions. The other images in the band serve for the purpose of defining one degree of freedom for which the energy of the climbing image is maximized along the band. In this work, we adopt the CINEB method to explore metastable CO states with distinct structural distortions following a computed structural path and compute the energy barrier along the path.
We obtain the structural path by defining two stable CO structures relaxed with different initial conditions and constructing an energy path between two structures using the CINEB method.

\subsection{Order Parameter}

While ferroelectricity is a phenomenon driven by the spontaneous polarization of materials, the polarization calculation in a periodic system requires a careful treatment of the formula~\cite{PhysRevB.48.4442}. At the same time, the inversion symmetry breaking of a structure is a clear indication of the spontaneous polarization. 
While we are not interested in obtaining the quantitative value of the polarization in this work, we will investigate the displacements of Fe and O planes in the rhombohedral unit cell along the $[111]_c$ direction (Figure \ref{1}), where the Fe plane distortion occurs below the critical temperature.

The displacements of Fe and O planes were investigated in the following way. First, we confirm that the CO1 and CO3 structures are centrosymmetric and define the central plane $C$ as the midway between
Fe1 and Fe6 planes. 
Next, we generated the other dashed-line planes which are equidistant and correspond to the Fe and O planes of the undistorted high-temperature structure (see Figure \ref{2}). Then, we can quantify how much Fe and O planes are displaced from the dashed lines. We define the total displacements per unit cell ($\Delta_{tot}$) for these Fe and O planes (see Table \ref{structural parameters}). 
For the CO2 structure, the $\Delta_{tot}$ is finite due to the inversion symmetry breaking, also implying the emergence of ferroelectricity.

\section{Results and Discussions}

\subsection{Structural Relaxations}

\begin{figure}[htb!]
  \includegraphics[width=0.40\textwidth]{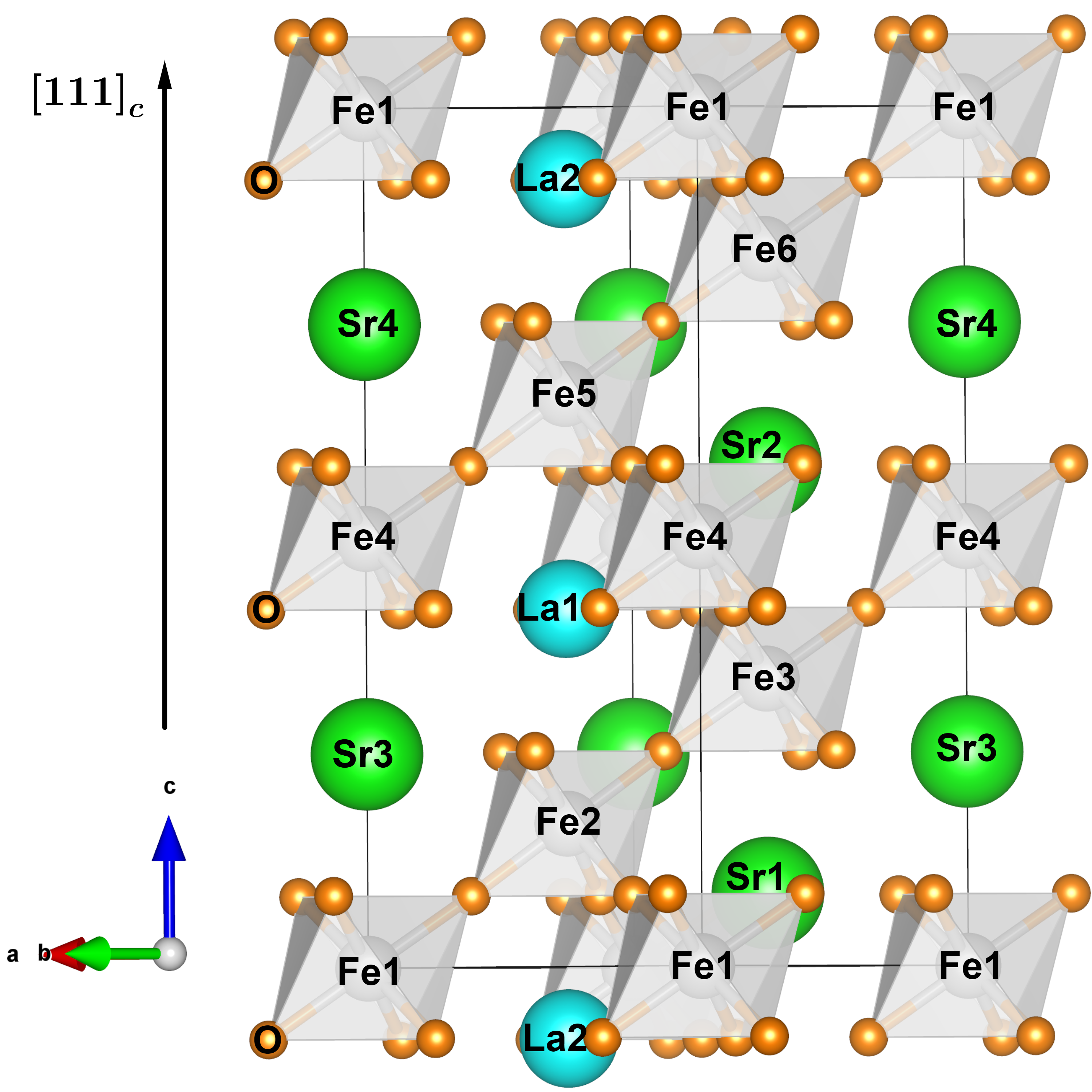}
\caption{The crystal structure of La$_{1/3}$Sr$_{2/3}$FeO$_3$ in a rhombohedral unit cell. The rhombohedral c-axis is equivalent to the $[111]_c$ direction of the cubic unit cell.
}
   \label{1}
\end{figure}

The study of the neutron diffraction measurement by Battle \textit{et al}~\cite{Battle1} at the room temperature showed that bulk LSFO forms a rhombohedral structure of the space group $R\bar{3}c$ (Figure \ref{1}) with lattice constants of $a=5.47$ \AA\: and $c=13.35$ \AA \:. The rhombohedral unit cell of LSFO has 30 atoms including two La, four Sr, six Fe, and eighteen O ions. The c-axis of the rhomboheral unit cell is equivalent to the $[111]_c$ direction of the pseudocubic one, which is conventionally adopted in literatures. Thus, the cubic $[111]_c$ direction will be also adopted in this paper. 
Above the CO critical temperature, all Fe ions are equivalent and have the same Fe-O bond lengths.
As the temperature is lowered below T$_{CO}$, both SDW and CDW orders develop along the [111]$_c$ direction. While the CDW order spans the periodicity of three Fe ions, the antiferromagnetic SDW  repeats in the unit-cell of six Fe ions, which is commensurate with the crystal lattice periodicity.
As a result, the space group of the crystal structure is lowered to $P\bar{3}m1$ 
(trigonal, No. 164) with the point group symmetry of $D_{3d}$
while the crystal remains centrosymmetric.
Here, we find that three distinct CO phases can be stable in LSFO with the same commensurate modulations of the SDW and CDW, and the stabilities of these CO structures are dependent on the electronic correlation effect (the Hubbard $U$ values).

\begin{figure}[htb!]
  \includegraphics[width=0.48\textwidth]{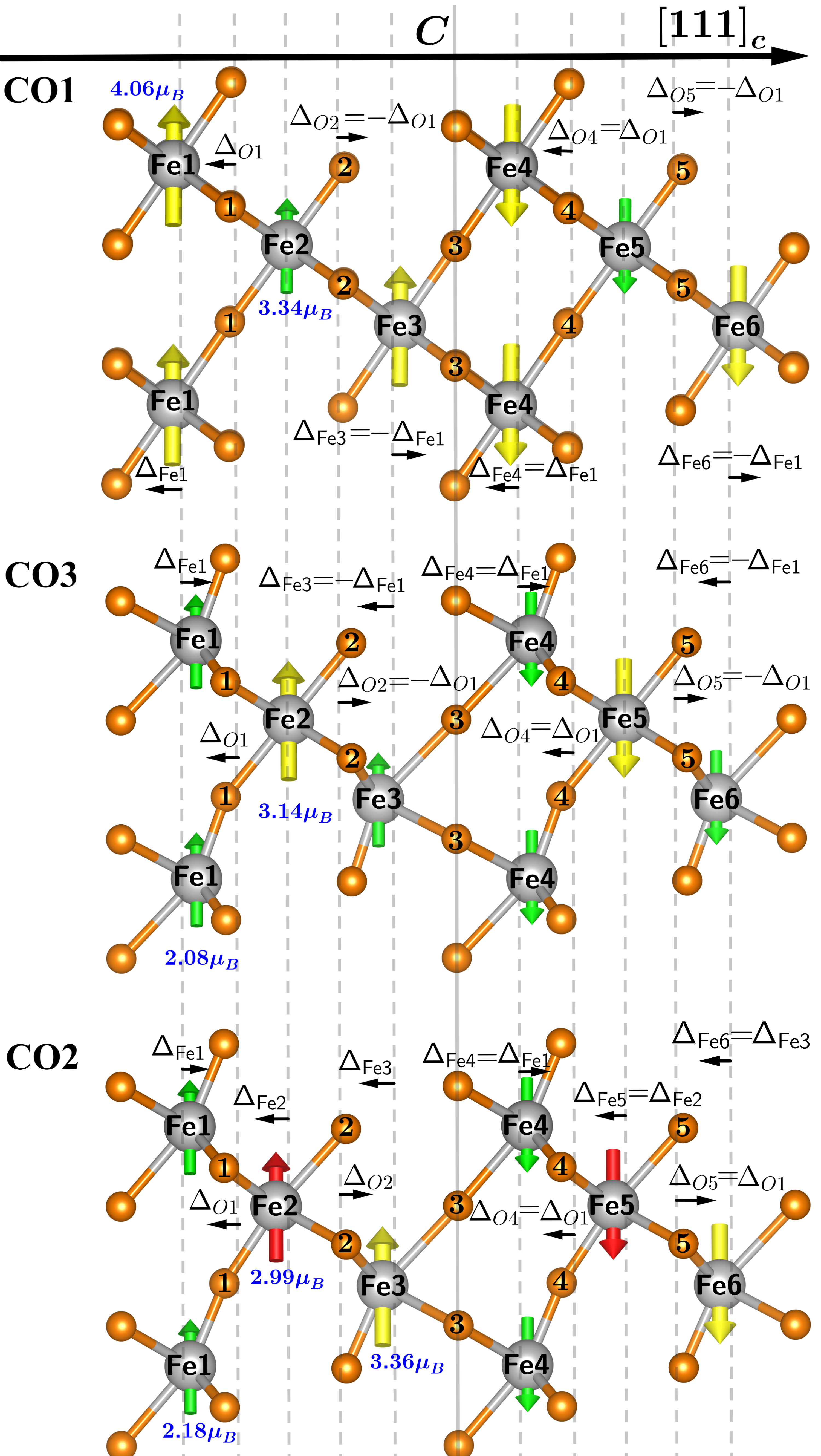}
\caption{The schemetics of Fe magnetic moments and the displacements of Fe/O planes for different CO1, CO2, and CO3 phases along the $[111]_c$ direction. 
The displacements are the changes of atomic positions from their undistorted structures (grey dash lines). The central plane C (grey solid line) is midway between Fe1 and Fe6 planes. 
}
   \label{2}
\end{figure}

\begin{table}[htb!]
\caption{The relaxed cell parameters ($a$ and $c$), the space group, 
the Fe plane displacements, and the total displacement ($\Delta_{tot}$) of LSFO at each CO phase. Both CO1 and CO3 structures (space group: $P\bar{3}m1$) have the inversion symmetry, while CO2 and CO$\bar{2}$ structures (space group: $P3m1$) do not.}
\begin{tabular}{c|c|c|c|c}
\hline
CO phase   
&c (\AA) & a (\AA)
& $\Delta_{Fe1}/\Delta_{Fe3}$/$\Delta_{Fe2}$ [\AA]  
& $\Delta_{tot}$ [\AA]

  \\ [0.5ex] 
\hline\hline
CO1 
& 13.12 & 5.38
&  -0.01/0.01/0.00 
& 0.00
\\
CO3 
&13.22&5.36
& 0.03/-0.03/0.00 
& 0.00
\\
\hline
CO2  
&13.19& 5.37 
& 0.01/-0.04/-0.03 
&-0.22
\\
CO$\bar{2}$ 
&13.19 & 5.37
&0.04/-0.01/0.03 
&0.22
\\
\hline
\end{tabular}
\label{structural parameters}
\end{table}

As already noted from Ref.\:\onlinecite{Ultrafast}, two distinct centrosymmetric CO structures 
(CO1 and CO3, see Fig.\:\ref{2}) can be obtained by relaxing them with different Hubbard $U$ values in DFT$+U$.
To explore other metastable CO phases and the energy barriers, these two CO1 and CO3 structures will be used for two reference structures in the CINEB method.
The first stable structure CO1 (charge ordering 1) was obtained in a strongly correlated regime with $U$=5 eV and $J$=1 eV, which have been used for LSFO in literatures \cite{DFT+U,Ultrafast,Ferro}. Another stable structure CO3 was obtained in a weakly correlated regime \cite{DFT+U} with $U$=3 eV and $J$=0.6 eV, which was also used by Zhu \textit{et al}~\cite{Ultrafast}. Both CO1 and CO3 structures exhibit a sixfold (six Fe ions) spin density wave (SDW) along the cubic [111]$_c$ direction, such that Fe1$(\uparrow)$Fe2$(\uparrow)$Fe3$(\uparrow)$Fe4$(\downarrow)$Fe5$(\downarrow)$Fe6$(\downarrow)$ (Figure \ref{2}).

We find that the $<$Fe-O$>$ mean bond-lengths are closely related to the magnetic moments. In the CO1 structure, the magnetic moments of Fe1(Fe4) and Fe3(Fe6) are larger than the one of Fe2(Fe5), so the charge states of Fe1 and Fe3 should be larger than the one of Fe2 (see Section \:\ref{sec:Correlation}). As a result, the high-spin (HS) Fe-O bond expands and the bond-disproportionation occurs. Particularly, in the CO1 structure the $<$Fe-O$>$ mean bond-lengths of Fe1-Fe2-Fe3 (similar to Fe4-Fe5-Fe6) show the big-small-big pattern, coupled to the big-small-big magnetic moments of Fe1, Fe2, and Fe3 ions respectively (Fig.\:\ref{2}). The $<$Fe-O$>$ mean bond-lengths of CO1 are 1.92\:\AA\: for Fe1 and 1.86\:\AA\: for Fe2, 
respectively.

In the case of CO3, the magnetic moments of Fe1 (Fe4) and Fe3 (Fe6) ions become smaller than Fe2 (Fe5) ion and the bond-length changes to small-big-small.
The bond-lengths are 1.88\:\AA\: for Fe1 and 1.94\:\AA\: for Fe2.
As a result of the Fe-O bond-length disproportionation,
the displacements of Fe and O planes are also non-uniform as shown in Fig.\:\ref{2}.

In Table \ref{structural parameters}, we list the relaxed unit-cell parameters, the space group, the displacements of Fe planes, and the total displacement of Fe and O planes ($\Delta_{tot}$) along the [111]$_c$. Both of CO1 and CO3 have the space group of $P\bar{3}m1$, implying CO1 and CO3 are centrosymmetric. Also, based on the displacements of Fe and O planes of CO1 and CO3, the Figure \ref{2} shows that the reflection of the supercells, including Fe1O$_6$-Fe6O$_6$ cells, of CO1 and CO3  around the central plane C yields the same supercells respectively. Finally, the total displacements of Fe and O planes in CO1 and CO3 are also zeros (Table \ref{structural parameters}), \textcolor{blue}{and their charge and magnetic orderings also centrosymmetric}, meaning polarization is not induced in CO1 and CO3.

\subsection{Energy Path along Multiple Charge Orderings}
\label{sec:path}

\begin{figure}[htb!]
  \centering
  \includegraphics[width=0.9\linewidth]{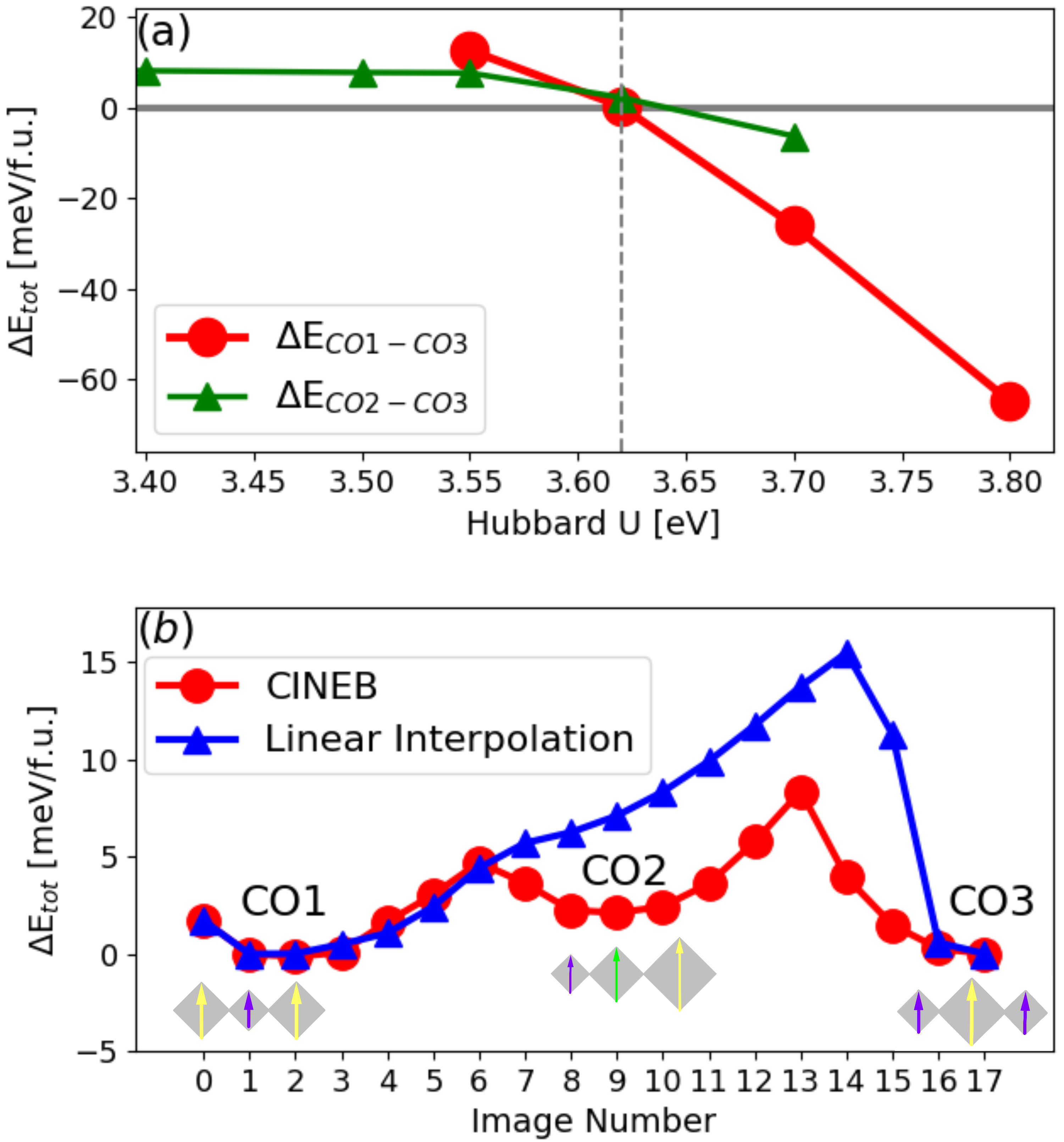}
    \caption{(a) The relative energies per formula unit of CO1, CO2, and CO3 phases as a function of the Hubbard $U$. (b) Comparison of CINEB and Linear Interpolation energies vs Image structures calculated with DFT+U at $U$=3.62 eV and $J$=0.724 eV.}
  \label{EnergyvsU}
\end{figure}

In this section, we compute the energy path between two energetically degenerate CO phases (CO1 and CO3) to explore the possible metal-stable CO states along the structural path.
We first tuned $U$ ($J$=0.2$U$) for CO1 and CO3 phases while relaxing crystal structures at a fixed volume to investigate their stability and plot the relative energy 
$\Delta$E$_{CO1-CO3} ( = E[\mathrm{CO1}] - E[\mathrm{CO3}])$
between them as a function of $U$ in Fig.\:\ref{EnergyvsU}(a). 
Here, we find that the low-temperature experimental ground-state structure (CO1) is stable when the $U$ value becomes larger than 3.7 eV (see Fig. \ref{BdvsU}(a)). 
While DFT$+U$ is a zero-temperature theory, we find that the energetics of different CO phases can be tuned by \textcolor{blue}{reducing} the onsite Coulomb repulsion $U$, which can mimic the effect of \textcolor{blue}{raising} temperature, \textcolor{blue}{applying}  pressure \textcolor{blue}{in experiments}, or photoinduced excitation. 
In principle, laser excitation in experiment can modulate the electronic structure from the ground-state, 
by affecting the exchange interaction \cite{Mentink}, and may eventually trigger a phenomenon called ``photoinduced structural phase transition" \cite{Nasu}.

Fig.\:\ref{EnergyvsU}(a) shows that the energy difference between CO1 and CO3 structures can be almost zero at $U_c=3.62$ eV ($J=0.724$ eV). 
This means that other meta-stable CO structures could be found in the CINEB calculation near $U=3.62$eV.
At $U$=3.62 eV, CO1(CO3) still has the big-small-big (small-big-small) bond-order (see Fig.\:\ref{BdvsU}).
Here, we perform a CINEB calculation at $U=3.62$ eV using both CO1 and CO3 as two reference structures. 
Remarkably, Fig.\:\ref{EnergyvsU}(b) shows that the CINEB energy curve calculated with $U_c$=3.62 eV can capture a meta-stable structure of CO2, whose energy is only 3meV above the CO1 or CO3 structure with the energy barrier of $\sim$7meV. 
This CO2 structure is obtained by the spontaneous displacement of the Fe plane and it can not be captured by the linear interpolation method where the image structures along the path are obtained by linearly interpolating atomic positions between CO1 and CO3.

The obtained CO2 structure has the small-medium-big FeO$_6$ octahedra (Fe-O bond order), 
(see Fig.\:\ref{BdvsU}(b)), coupled to the magnetic ordering of 2.2 (small), 3.0 (medium), and 3.4 $\mu_B$ (big). 
Unlike CO1 and CO3, CO2 has the space group of $P3m1$ 
(trigonal, No. 156) 
with the point group symmetry of $C_{3v}$ , which belongs to a polar point group \cite{Ferro_Point_Group}. 
The reflection of CO2 supercell around the C plane does not yield the same supercell (see Fig.\:\ref{2}), implying the broken inversion symmetry in CO2. Also, the total displacement ($\Delta_{tot}$) of Fe and O planes in CO2 is not zero (see Table \ref{structural parameters}), resulting \textcolor{blue}{a polar distortion in the structure}. Remarkably, We also find that the other meta-stable CO$\bar{2}$ structure can be obtained by applying the inversion operation to the CO2 structure about the central plane in Fig.\:\ref{2} and shows the opposite polarization compared to the CO2 case (see Table \ref{structural parameters}). \textcolor{blue}{Since CO2 and CO$\bar{2}$ are metallic (see Sec.\:\ref{sec:Correlation} ), equivalent in energy, and inversion of each other, they are polar metals\cite{CriteriaFerroelectrics1,CriteriaFerroelectrics2}}.

\begin{figure}[htb!]
  \includegraphics[width=0.95\linewidth]{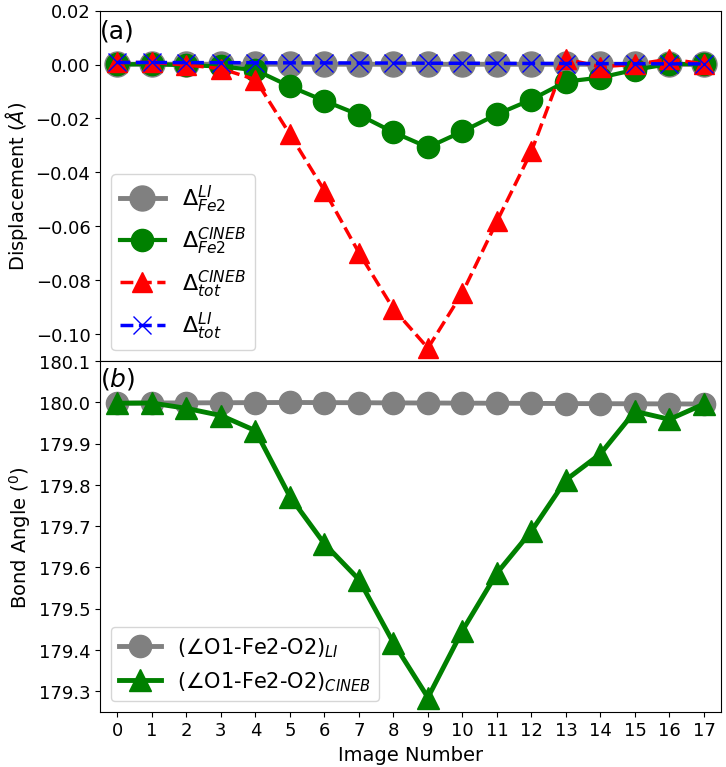}
\caption{ Comparison between the CINEB and the Linear Interpolation paths: (a) The displacement of Fe2 plane and the total displacement of Fe and O planes. (b) The bond angle $\angle$O1-Fe2-O2.}
   \label{5}
\end{figure}

To address the difference between the CINEB and Linear Interpolation results, we compare
the displacement ($\Delta_{Fe2}$) of Fe2 plane, the total displacement ($\Delta_{tot}$) of Fe and O planes, and the bond angle along O1-Fe2-O2 ($\angle$O1-Fe2-O2).
Figure \ref{5}(a) shows that along the Linear Interpolation path the displacement of Fe2 plane $\Delta_{Fe2}^{LI}$ and the total displacement of Fe and O planes $\Delta_{tot}^{LI}$ are kept zero, while along the CINEB path an abrupt change of the displacement of Fe2 plane $\Delta_{Fe2}^{CINEB}$ and the total displacement $\Delta_{tot}^{CINEB}$ occurs at image number 5 and reach a minimum at image 9 where CO2 was captured. This change of the Fe2 displacement along the CINEB path is also accompanied by a sudden change of the bond angle ($\angle$O1-Fe2-O2)$_{CINEB}$, while the one along the Linear Interpolation path ($\angle$O1-Fe2-O2)$_{LI}$ remains $180^0$ (Figure \ref{5}(b)). The existences of this CO2 phase might be captured by Sabyasachi \textit{et al.} and Yang \textit{et al.}, where the neutron diffraction shows the multiple $Q$ plane magnetic reflections with equivalent intensities~\cite{M&X&N1,M&X&N3}.

\subsection{The dependence of structural parameters on $U$}
\label{sec:Udep}
\begin{figure}[htb!]
  \centering
  \includegraphics[width=0.9\linewidth]{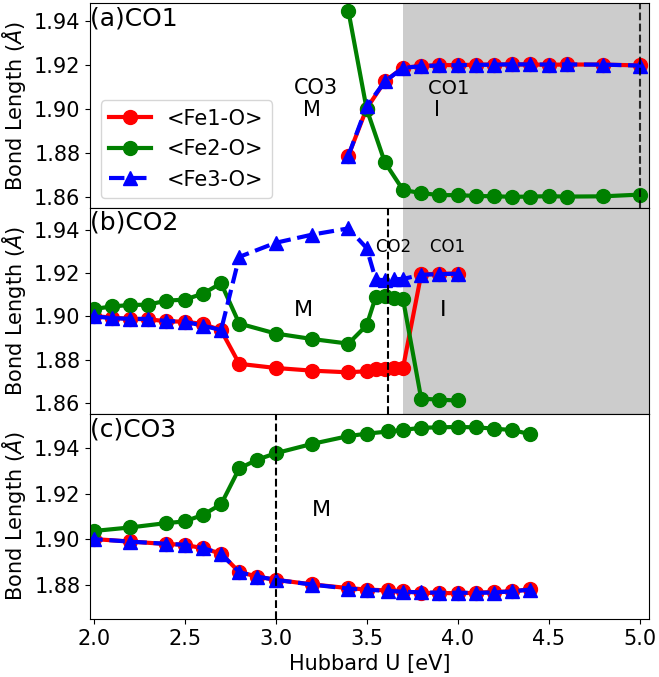}
    \caption{The $<$Fe-O$>$ mean bond lengths vs the Hubbard $U$ ($J=0.2U$)  obtained for (a) CO1, (b) CO2, and (c) CO3 phases. The vertical dash lines represent the $U$ values used for stabilizing each CO phase, namely CO1 ($U=5.0$ eV), CO2 ($U=3.62$ eV), and CO3 ($U=3.0$ eV) as shown in Table \ref{structural parameters}. The white (shaded) regions represent metallic (insulating) phases.
    }
  \label{BdvsU}
\end{figure}

Our structural relaxation results show that the stability of the different CO phases depends sensitively on the Hubbard $U$ values ($J=0.2U$).
In general, the correlation effect of the Hubbard $U$ is important to stabilize the bond/charge disproportionation in many oxides including nickelates~\cite{PhysRevLett.109.156402,PhysRevB.103.085110}, cobaltates~\cite{PhysRevB.101.195125}, ferrites~\cite{PhysRevX.8.031059}, and manganites~\cite{PhysRevB.92.115143}.
This is because only parts of the transition metal $M$ sites undergo the spin-state transition to the HS state with the $M-O$ bond elongation and more HS sites are populated with the stronger $U$ values.
Our calculation confirms that the DFT-relaxed structure of LSFO shows no Fe-O bond disproportionation, consistently as the experimental high-temperature structure, and the increase of $U$ energetically favors the structures with more HS states in a non-trivial way.

Fig.\:\ref{BdvsU}(a) shows that the CO1 structure as shown in Table \ref{structural parameters} can be stable only when $U> 3.7$ eV ($J$=0.74 eV) and the structural transition to CO3 occurs along with the insulator-metal transition. The CO2 structure as shown in Table \ref{structural parameters} is meta-stable in a narrow $U$ range of  $3.55\leq U \leq 3.7$ eV and evolves into a distorted CO3 phase as $U$ becomes lower than 3.55 eV. The CO3 structure can be stable in a wide-range of $U$ values although this phase is energetically lower than CO1 or CO2 phases when $U \leq 3.62$ eV. Both of CO2 and CO3 structures converge to the high-temperature structure without the Fe-O disproportionation as $U$ becomes smaller than 2eV. 
We find that the insulating phase in LSFO occurs only in the CO1 structure with $U>$ 3.7eV.

\subsection{Electronic Structure and magnetism in LSFO}
\label{sec:Correlation}
Here, we investigate electronic structures of LSFO at different CO states computed using DFT+U.
Due to the AFM structure, the Fe1/Fe2/Fe3 density of states (DOS) is equivalent to the Fe4/Fe5/Fe6 one in LSFO once their spins are flipped. 
For CO1 and CO3, the crystal structures are centrosymmetric and we show only Fe1 and Fe2 DOS since Fe1 (Fe4) and Fe3 (Fe6) are equivalent.
To distinguish the importance of electronic correlations from the structure effect, we compare $U=3.62$ eV ($J=0.724$ eV) and $U=4$ eV ($J=0.8$ eV) DOS at the fixed structure of each CO phase.

\begin{figure*}[htb!]
  \includegraphics[width=0.44\linewidth]{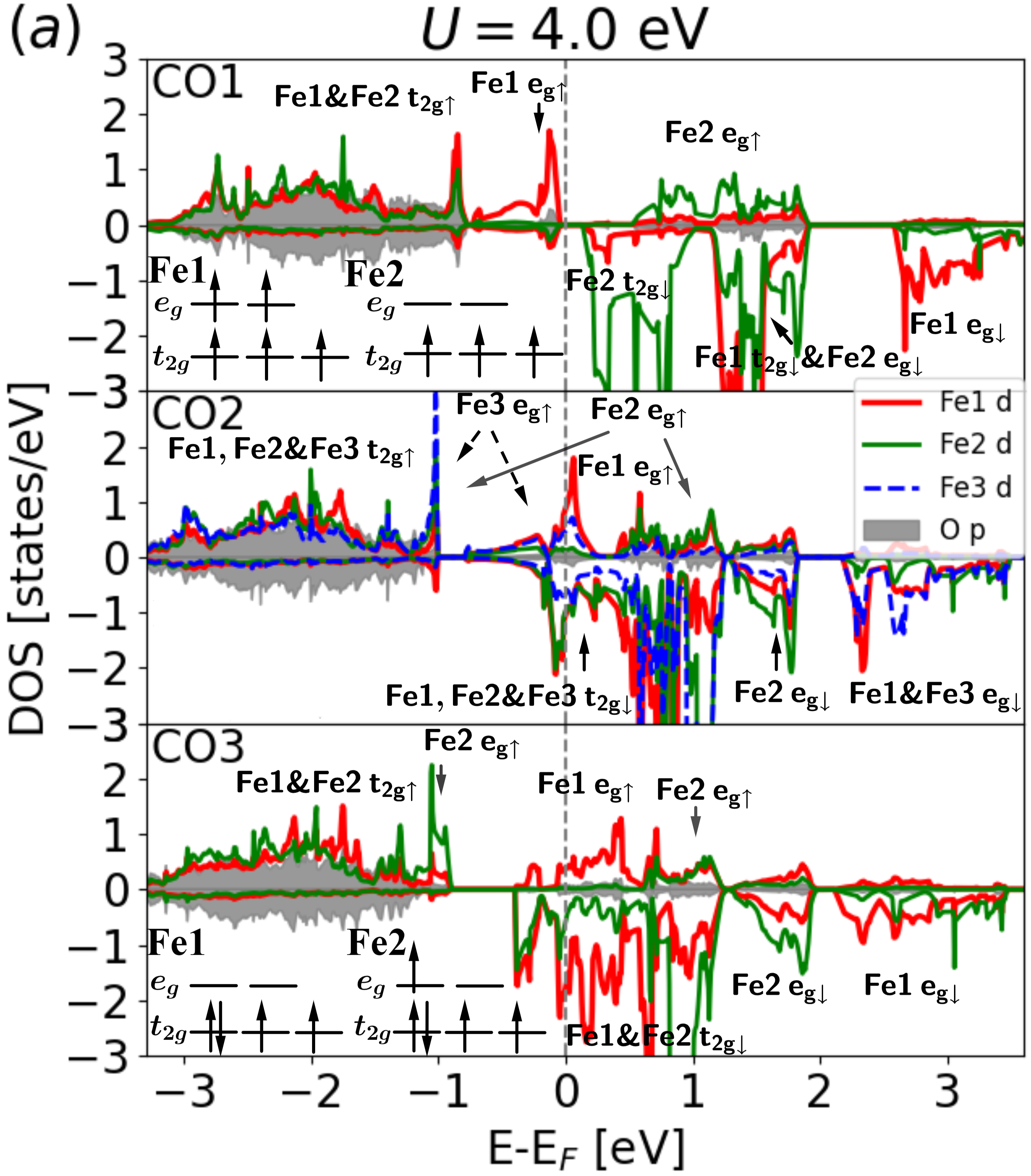}
  \includegraphics[width=0.44\linewidth]{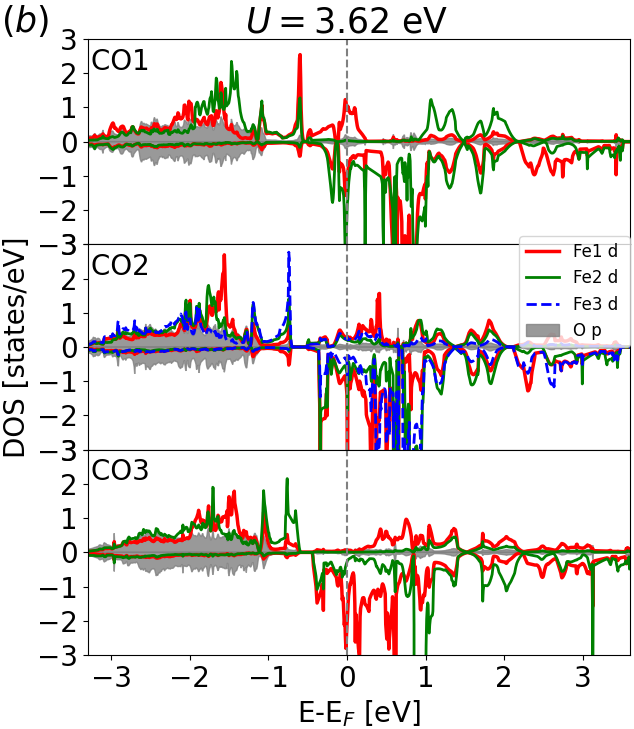}
  \caption{The DOS plots of CO1, CO2, and CO3 phases calculated with DFT$+U$. (a) $U=4.0$ eV and $J=0.8$ eV and (b) $U=3.62$ eV and $J=0.724$ eV. Schematic energy diagrams of Fe t$_{2g}$ and e$_{g}$ orbitals are also shown in the insets.
  }
  \label{6}
\end{figure*}

At $U$=4 eV, the CO1 phase is an insulating state with the spectral gap size of $\sim$120 meV (see Fig.\:\ref{6}(a)),  consistent with the optical gap measurement in LSFO at a low temperature \cite{Optical1}. 
In the Fe1 ion, both e$_g$ and t$_{2g}$ bands are half-filled with the gap size comparable to $U$ behaving as a typical Mott insulator. However, only the t$_{2g}$ bands of Fe2 are half-filled, while the e$_{g}$ bands are almost empty (see Fig.\:\ref{6}(a)). 
This is consistent with the high-spin picture of the charge-ordering state between Fe1 ($d^5$; t$_{2g\uparrow}^3$e$_{g\uparrow}^2$) and Fe2 ($d^3$; t$_{2g\uparrow}^3$e$_{g\uparrow}^0$) ions.
As the correlation becomes weaker ($U$=3.62 eV), the DOS for CO1 becomes metallic as the Fe1 e$_g$ (Fe2 t$_{2g}$) state is less (more) occupied and the spectral gap at the Fermi energy is closed.

In CO3 at $U$=4eV, the charge-ordering pattern changes for Fe1 ($d^4$; t$_{2g\uparrow}^3t_{2g\downarrow}^1e_{g\uparrow}^0$) and Fe2 ($d^5$; t$_{2g\uparrow}^3t_{2g\downarrow}^1e_{g\uparrow}^1$). The spin state for Fe1 changes to the low-spin, while the Fe2 spin is close to the intermediate one.
This is because the crystal field splitting of the Fe1 ion becomes larger due to the smaller octahedron size compared to Fe1 in the CO1 phase.
As a result, both Fe1 t$_{2g}$ and Fe2 t$_{2g}$ states are partially filled and the DOS becomes metallic (see Fig.\:\ref{6}(a)).
As the correlation becomes weak ($U$=3.62 eV), the CO3 phase remains metallic.

Similar to CO3, CO2 is metallic at both $U$=4eV and 3.62eV. 
As the Fe1 $d$ DOS of CO2 is similar to the Fe1 $d$ DOS of CO3 and the Fe1-O bond-lengths of CO2 and CO3 are similar to each other as well, we expect that the local electronic configuration of Fe1 should be similarly given as the low-spin $d^4$ (t$_{2g\uparrow}^3t_{2g\downarrow}^1e_{g\uparrow}^0$).
Moreover, the Fe-O bond-length of the Fe1 ion is the smallest, while those of Fe2 and Fe3 ions are close to each other implying the similar electronic structure between Fe2 and Fe3. 
Nevertheless, the evidence of the CO can be found near $E\approx -1 eV$ where the occupied Fe3 e$_{g}$ states have slightly more DOS than the Fe2 one, while their t$_{2g}$ DOS are similar. This implies that the local electronic configurations of Fe2 and Fe3 ions should be Fe$^{(3.5+\delta)+}$ (t$_{2g\uparrow}^3t_{2g\downarrow}^1e_{g\uparrow}^{0.5-\delta}$) and Fe$^{(3.5-\delta)+}$ (t$_{2g\uparrow}^3t_{2g\downarrow}^1e_{g\uparrow}^{0.5+\delta}$), respectively.

The calculated magnetic moments of Fe ions ($m_{Fe1}$, $m_{Fe2}$, and $m_{Fe3}$) are coupled to the above valence states and these values for CO1, CO2, and CO3 are shown in Table \ref{Magnetic_Moment}. The magnetic moments in CO1 calculated with DFT+U ($U$=4 eV, $J$=0.8 eV) are in a good agreement with the experimental ones recently obtained by Li \textit{et al}~\cite{MSandND} at low temperature.
The calculated value of $m_{Fe1}$ is rather screened from the electronic configuration estimation based on the DOS since we expect $m_{Fe1}$ (t$_{2g\uparrow}^3$e$_{g\uparrow}^2$) = 5$\mu_B$, while the $m_{Fe2}$ value is consistent (t$_{2g\uparrow}^3$e$_{g\uparrow}^0$ = 3$\mu_B$).
The expected moments of the Fe1 and Fe2 ions in CO3 are $m_{Fe1}$ (t$_{2g\uparrow}^3$t$_{2g\downarrow}^1$e$_{g\uparrow}^0$) = 2$\mu_B$ and $m_{Fe2}$ (t$_{2g\uparrow}^3$t$_{2g\downarrow}^1$e$_{g\uparrow}^1$) = 3$\mu_B$, respectively. However, since CO3 is metallic and the magnetic moments calculated with DFT+$U$ also depends on $U$ and $J$, our calculated moments of  2.42$\mu_B$ and 3.52$\mu_B$ at $U$=4 eV are larger than these expected values. We confirmed that the magnetic moments of Fe1 and Fe2 are reduced at $U$=3 eV as 2.08 and 3.14 $\mu_B$, similar to the expected values respectively.

Similarly, the magnetic moment of Fe1 at CO2 calculated with $U$=4 eV is 2.70 $\mu_B$, which is still large for a LS state of Fe$^{4+}$ (2.0 $\mu_B$).
We find that this moment computed using $U$=3.62eV is more consistent as 2.18$\mu_B$.
For CO2, the magnetic moments of $m_{Fe1}$, $m_{Fe2}$, and $m_{Fe3}$ show the small-medium-big pattern, which is consistent with the charge-ordering pattern of  Fe$^{4+}$-Fe$^{(3.5+\delta)+}$-Fe$^{(3.5-\delta)+}$.

\begin{table}[htb!]
\caption{Magnetic moments of Fe1, Fe2, and Fe3 ions, namely $m_{Fe1}$, $m_{Fe2} $, and $m_{Fe3}$ calculated with DFT$+U$ and compared with the experimental ones obtained by Li \textit{et al}\cite{MSandND}}
\resizebox{8.5cm}{!}{
\begin{tabular}{c|c|c|c}
\hline
LSFO    &$m_{Fe1} [\mu_B]$  & $m_{Fe2} [\mu_B]$  & $m_{Fe3} [\mu_B]$   \\ [0.5ex] 
\hline
CO1 (U=4.0 eV)  & 3.74 (Fe$^{3+}$)  & 3.12 (Fe$^{5+}$) & 3.74 (Fe$^{3+}$) 
\\ 
Experiment.\cite{MSandND} & 3.67  (Fe$^{3+}$) &  3.26 (Fe$^{5+}$) & 3.67 (Fe$^{3+}$)
\\
CO3 (U=4.0 eV) & 2.42 (Fe$^{4+}$) & 3.52 (Fe$^{3+}$) & 2.42 (Fe$^{4+}$)
\\
CO2 (U=4.0 eV) & 2.70 (Fe$^{4+}$) & 3.26 (Fe$^{(3.5+\delta)+}$) & 3.60 (Fe$^{(3.5-\delta)+}$)
\\
CO2 (U=3.62 eV) & 2.18 (Fe$^{4+}$) & 2.99 (Fe$^{(3.5+\delta)+}$) & 3.36 (Fe$^{(3.5-\delta)+}$)
\\
\hline
\end{tabular}}
\label{Magnetic_Moment}
\end{table}

\section{Conclusion}

In conclusion, we studied the structural and electronic properties of charge-ordered La doped SrFeO$_3$,  La$_{1/3}$Sr$_{2/3}$FeO$_{3}$ (LSFO) systematically using DFT+U along with the antiferromagnetic order.
We find that metastable structures with distinct CO phases in LSFO can be obtained by relaxing the structures with the different $U$ values varying the correlation effect.
The DFT+U calculation of LSFO with $U$=5eV can capture the low temperature CO phase (CO1 in the main text) of the big-small-big pattern, where the enhanced charge density is accompanied by the large magnetic moment with the Fe-O bond elongation.
The ground-state is insulating as the spectral function at the Fermi energy opens a Mott gap driven by the high-spin states of Fe ions.

As the correlation effect becomes weak by reducing the $U$ value in DFT+U, we can capture other metastable CO phases with distinct Fe-O bond patterns.
One CO phase (CO3 in the main text) shows the crystal structure with the same space group as the CO1 phase, while the CO pattern changes to small-big-small.
The other metastable CO phase (CO2 in the main text) can be obtained by interpolating the structural path between CO1 and CO3 phases using the CINEB calculation.
Remarkably, the CO2 phase stabilize a lower symmetry crystal structure along with the inversion symmetry breaking and it shows the \textcolor{blue}{polar distorted structure} driven by the big-medium-small CO pattern.
This CO2 phase can not be captured by the linear interpolation method as it requires the spontaneous displacement of Fe ions at the symmetric points.
The electronic structures of these metastable CO states are notably changed as both CO2 and CO3 phases are metallic while the ground-state CO1 phase is insulating.
The energy barrier of this CO2 phase along the structural path is only $\sim$7meV.

Our results suggest that the strong correlation effect plays an important role to study and stabilize the multiple CO phases of transition metal oxides accompanying the mixed valence and the metal-oxygen bond disproportionation.
The CINEB method combined with the energy and force calculations based on the first-principles can capture such metastable CO phases along with the distinct electronic structure from their ground state.
While DFT+U is an efficient static method to incorporate the correlation effect, it can generally suffer from the convergence problem in systems with multiple correlated states~\cite{PhysRevB.101.195125,Ratcliff_2021}.
More advanced first-principle method such as dynamical mean field theory (DMFT) can be a promising way to study metastable phases in strongly correlated materials driven by both the structural distortion and the strong correlations especially when the CINEB method is combined with the energy and force calculations within DMFT~\cite{PhysRevLett.112.146401,PhysRevB.94.195146}.

\section*{Acknowledgement}
We thank Yue Cao for fruitful discussions.
NN, AL, AN, and HP acknowledge financial support from the US Department of Energy, Office of Science, Office of Basic Energy Sciences, Materials Science and Engineering Division. VS was supported from NSF SI2-SSE Grant 1740112. We gratefully acknowledge the computing resources provided on Bebop, a high- performance computing cluster operated by the Laboratory Computing Resource Center at Argonne National Laboratory.


\bibliography{references.bib}


\appendix

\section{The comparison of DFT and DFT+U results}

Here, we compare both DFT ($U=J=0$) and DFT+U ($U=5$ eV \& $J=1$ eV) DoS plots for each fixed structure of CO1, CO2, and CO3 in Table \ref{structural parameters}.
\begin{figure*}[htb!]
  \centering
  \includegraphics[width=0.40\linewidth]{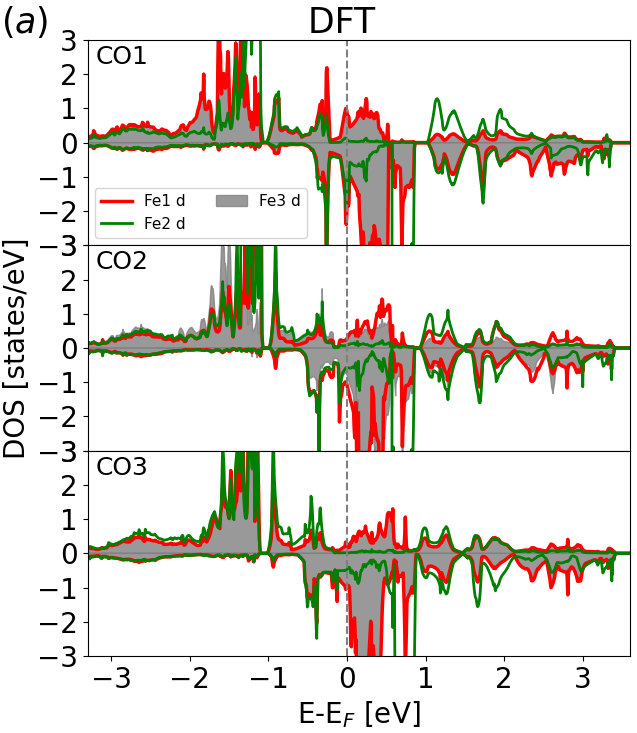}
  \includegraphics[width=0.40\linewidth]{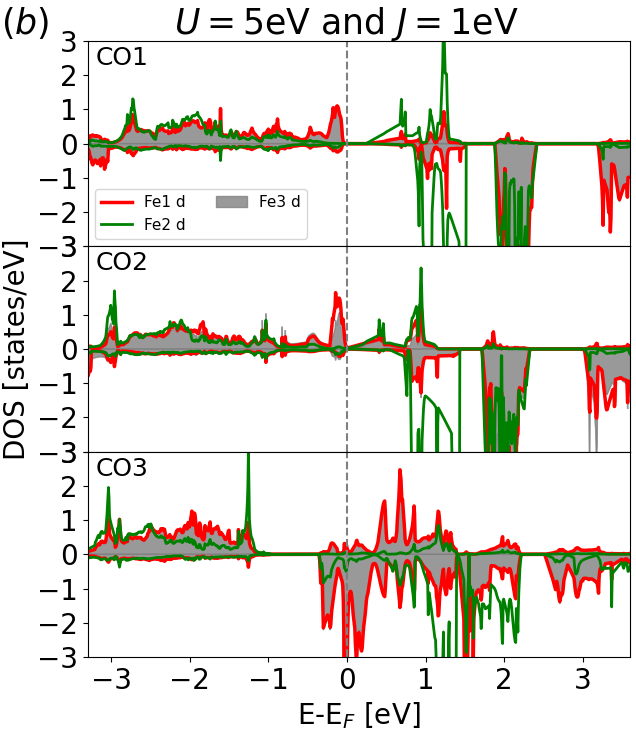} 
  \caption{DoS of the fixed structures CO1, CO2, and CO3, calculated with a) $U=J=0$ and b) $U=5$ eV \& $J=1$ eV. 
  }
  \label{DFTvsDFT+U}
\end{figure*}
We find that the charge ordering pattern, which is dictated by the pattern of magnetic moments in Table \ref{Mag_DFT_vs_DFT+U}, remains the same (CO1 (big-small-big), CO2 (small-medium-big), and CO3 (small-big-small)) regardless of $U$ values, while the sizes of magnetic moments are reduced as the $U$ value decreases.
In DFT+U ($U=5$ eV \& $J=1$ eV), the spin-state transition can occur for CO2 since Fe1 changes from low spin Fe$^{4+}$ (t$_{2g\uparrow}^3$t$_{2g\downarrow}^1$e$_{g\uparrow}^0$) to high spin Fe$^{4+}$ (t$_{2g\uparrow}^3$e$_{g\uparrow}^1$), as also shown in the DoS plot for CO2 (Figure \ref{DFTvsDFT+U}(b)).
The DFT DoS plots for all CO structures (Fig. \ref{DFTvsDFT+U}(a)) show the metallic behavior with the smaller differences between Fe ions, while CO1 and CO2 are insulating in DFT+U (Fig. \ref{DFTvsDFT+U}(b)).
This implies that the metal-insulator transition will occur even for the fixed LSFO structure as the correlation ($U$) effect enhances the tendancy toward the charge/magnetic ordering.
\begin{table}[htb!]
\centering
\caption{Magnetic order as $m_{Fe1}$ $m_{Fe2} $ $m_{Fe3}$  of the fixed structures CO1, CO2, and CO3 calculated with U=J=0 and $U=5$ eV \& $J=1$ eV}
\resizebox{8.0cm}{!}{
\begin{tabular}{c|c|c}
\hline
   &\multicolumn{2}{c}{$m_{Fe1}$ $m_{Fe2}$ $m_{Fe3}$ $[\mu_B]$}  \\ [0.5ex] 
\hline
   & $U=J=0$ eV   & $U=5$eV \& $J=1$eV \\   
\hline
CO1  & 2.56 2.41 2.55   & 4.04 3.41 4.04   
\\ 
CO2 & 2.00 2.59 2.62    &  3.83 3.72 4.01   
\\
CO3 & 2.15 2.84 2.15  & 2.27 4.04 2.26 
 \\
\hline
\end{tabular}}
\label{Mag_DFT_vs_DFT+U}
\end{table}

\end{document}